 \documentclass[final,5p,times,twocolumn,authoryear]{elsarticle}

\usepackage{amssymb}
\usepackage{lipsum}
\usepackage{tabularx} 
\usepackage{graphicx} 
\usepackage{amsmath}

\newcommand{\gray}{$\gamma$-ray\ }
\newcommand{\grays}{$\gamma$-rays\ }
\journal{High Energy Astrophysics}

\begin{document}

\begin{frontmatter}



\title{Prospects for joint reconstruction of imaging air Cherenkov Telescope array and extensive air shower array}


\author[a,d]{Zhipeng Zhang}
\author[a,b,c,f]{Ruizhi Yang}
\author[d,e,f]{Shoushan Zhang}
\author[d,e,f]{LiQiao Yin}
\author[d,e,f]{Jiali Liu}
\author[d,e,f]{Yudong Wang}
\author[d,e,f]{Lingling Ma}
\author[d,e,f]{Zhen Cao}

\affiliation[a]{School of Astronomy and Space Science, University of Science and Technology of China, Hefei, Anhui 230026, China}
\affiliation[b]{CAS Key Laboratory for Research in Galaxies and Cosmology, Department of Astronomy, University of Science and Technology of China, Hefei, Anhui 230026, China}
\affiliation[c]{Deep Space  Exploration Laboratory School of Physical Sciences, University of Science and Technology of China, Hefei 230026, China}
\affiliation[d]{Key Laboratory of Particle Astrophysics, Institute of High Energy Physics, Beijing, China}
\affiliation[e]{Department of Physics, University of Chinese Academy of Sciences, Beijing, China}
\affiliation[f]{Tianfu Cosmic Ray Research Center, Chengdu, China}
\begin{abstract}
In this paper, we proposed a joint reconstruction of \gray events using both extensive air array (EAS) and Imaging air Cherenkov Telescope array (IACT). We considered eight Cherenkov telescopes to be built on the LHAASO (Large High Altitude Air Shower Observatory) site and investigate the improvement in differential sensitivity when combining the information from both IACT and Moun detectors of LHAASO-KM2A.  We found that due to the higher cosmic ray background rejection power and higher gamma ray retention ratio provided by muon detectors of LHAASO, such a joint reconstruction can significantly improve the sensitivity of IACTs, especially for extended sources and long exposure time. The sensitivity of the eight-telescopes array can be improved by  $25\% - 60\%$ in different energy ranges using joint reconstructions with muon detectors.
\end{abstract}




\end{frontmatter}




\section{Introduction}
\label{sec:intro}
Ground-based \gray astronomy hinges on the measurement of the cascade of secondary particles generated by the interactions of \grays with the Earth's atmosphere. This method is the only possible way due to the atmospheric opacity, which obstructs the direct observation of \grays. For \grays above $10~\rm MeV$ the pair productions dominate the \gray interaction, and the produced electron-positron pairs will further interact with the nuclei and magnetic fields in the atmosphere through bremsstrahlung and synchrotron processes which may also produce \grays. Such cascade processes will produce a large amount of secondary particles, which will reach the earth in a relatively short time span (of the scale of several nanoseconds) and be dubbed as air shower  \citep{gaisser90}. The ground-based \gray telescopes detect the secondaries in the air shower and reconstruct the direction and energy for the primary \gray. Generally speaking, there are two types of different ground-based \gray telescopes. One is extensive air shower (EAS) arrays which detect the secondary particles on the ground level directly using particle detectors, the other is the Imaging Atmospheric Cherenkov telescope array (IACTs) which detects the Cherenkov radiation of the relativistic secondary particle. The EAS array has a large field of view (FOV) and a very high duty cycle. But currently, the angular resolution of such array is limited, {typically worse than $0.2^{\circ}$}. IACTs, on the other hand, can achieve an angular resolution as good as several arcminutes using stereo reconstruction. However, {because IACTs detect visible light, they can only operate in clear night conditions}, resulting a much lower duty cycle. The optical system also requires a limited FOV to sustain a reasonable imaging quality. As a result, the EAS array is extremely suitable for sky survey thanks to its high duty cycle and large FOV, while IACTs are good at investigating detailed structures due to its better angular resolution. In this regard, both instruments can complement each other. 

The existing IACT array include H.E.S.S  \citep{aharonian2006observations}, MAGIC  \citep{aleksic2012performance}, and VERITAS  \citep{kieda2013gamma}. while As$\gamma$  \citep{amenomori1999observation}, ARGO-YBJ  \citep{bartoli2015crab} and HAWC  \citep{abeysekara2017observation} represent the operating EAS arrays.  During the last three decades, these instruments have discovered more than 200 sources in the TeV band,  advancing the development of very-high-energy gamma-ray astronomy. {The next generation of IACTs, such as the Cherenkov Telescope Array (CTA) \citep{federici2011design} and ASTRI \citep{scuderi2022astri}, will further advance research in this area.}

As one of the latest EAS array, the Large High Altitude Air Shower Observation (LHAASO)  \citep{cao2019large}, located at Haizi Mountain, Daocheng, Sichuan province, China, is composed of three subarrays, including the $1{\rm km}^2$ array (LHAASO-KM2A), the water Cherenkov detector array (LHAASO-WCDA), and the wide field air Cherenkov/fluorescence telescopes array (LHAASO-WFCTA). LHAASO-KM2A consists of 5195 scintillator counters as electromagnetic particle detectors (ED) on the surface and 1188 underground muon detectors (MD) spread out
over a circular area of $1.3~\rm km^2$. Thanks to its large collection area and effective particle distinguish ability offered by MDs, LHAASO-KM2A can provide unprecedented \gray detection sensitivities above 20 TeV  \citep{aharonian2021observation}. On the other hand, because of the limited effective area $\sim 10^5 \rm~{m^2}$ of the existing IACTs, no source has been firmly detected by IACTs in ultra high energy domain($> 0.1~\rm PeV$) 
  \citep{cao2021ultrahigh}. Thus the current IACTs cannot provide sufficient synergy with LHAASO.  The next generation IACT array CTA-North has a similar latitude with LHAASO and can have a kind of synergy with LHAASO, but they have different science objectives.{ CTA-North will focus more on low energy observations, while LHAASO is more concerned with much higher energy sources, such as PeVatrons.} So in order to advance in this direction, the IACT Project Large Array of  Imaging Atmospheric Cherenkov Telescope arrays (LACT) in LHAASO site was proposed to take advantage of better angular resolution and have a good combination with LHAASO excellent sensitivity  \citep{Zhang2023pos}. LACT will have a common science goal with LHAASO and mainly focus on energy above  $10~ \rm TeV $. At the same time, utilizing the WhiteRabbit time synchronization \citep{cao2019large}, we have the same shower simultaneously observed both by LHAASO-KM2A and LACT, the information obtained from LHAASO-KM2A can be leveraged to enhance the performance of the IACTs.

The paper is organized as follows: In section2, we introduce the  IACT array configurations, which include the layout of the cherenkov telescope and the telescope parameters used in the simulation. In section3, we will discuss the generation of {monte carlo simulation of} extensive air shower and related parameters and discuss the reconstruction methods.  In section4, we discuss our results and possible implications. In the last, We also do a brief introduction to the LACT project.

\section{IACT Array Configuration}
\begin{figure}
  \centering
  \includegraphics[width=\columnwidth]{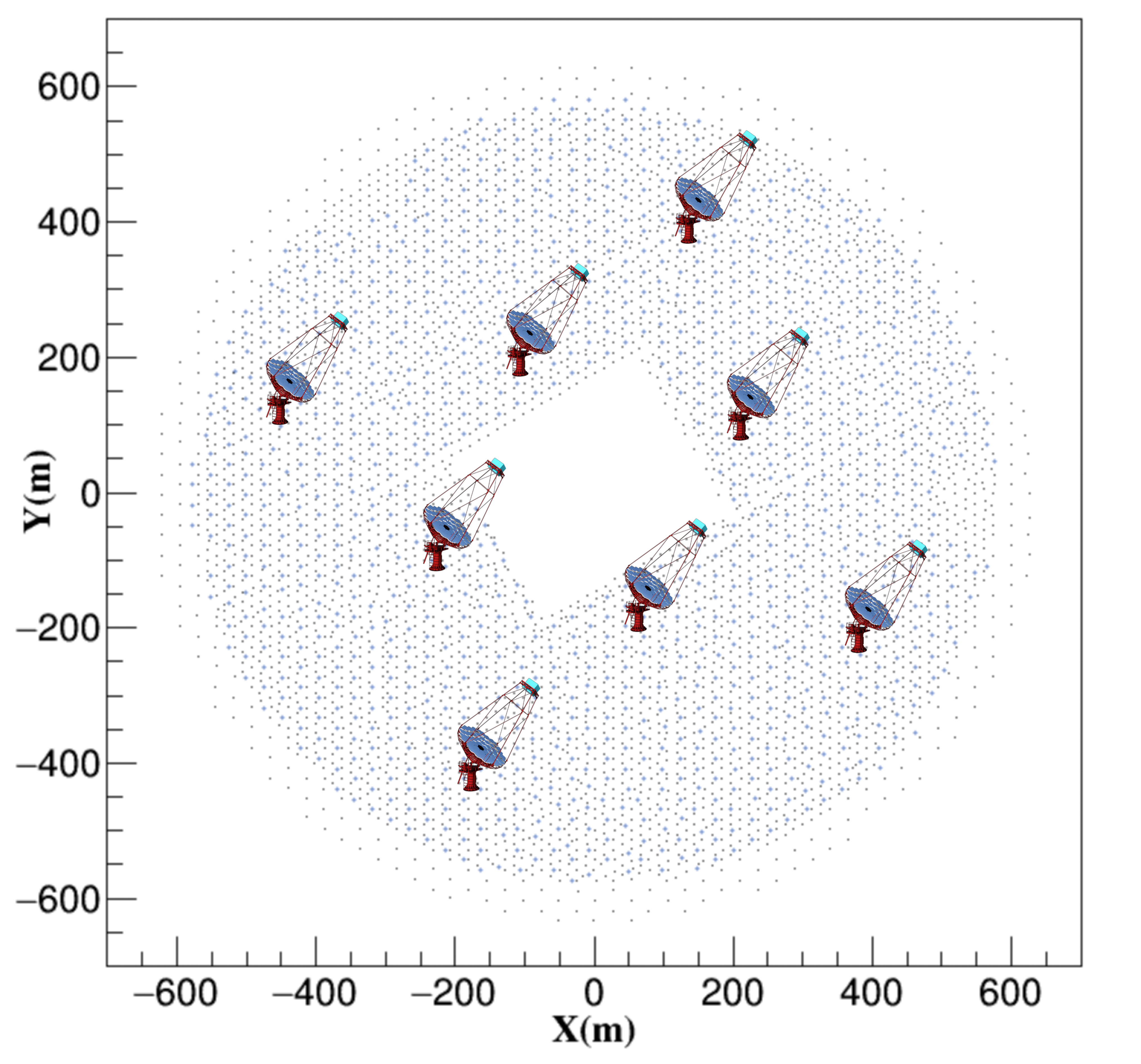} 
  \caption{Layout of the LHAASO and IACT array. {The blue squares represent muon detectors, while the black dots represent electromagnetic particle detectors.}}
  \label{fig:layout}
\end{figure}

\begin{figure}
  \centering
  \includegraphics[width=\columnwidth]{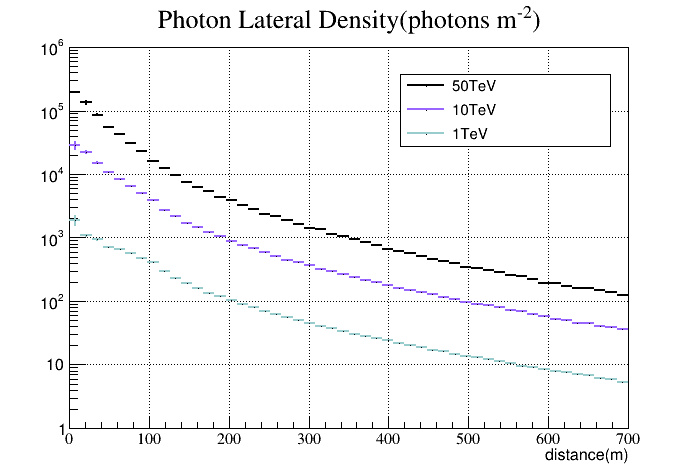} 
  \caption{Lateral photon density for 1 \rm TeV,10 \rm TeV and 50 \rm TeV gamma shower in 4400m altitude. The Data is obtained from Coriksa(IACT) and does not consider the atmosphere absorption.}
  \label{fig:photon_density}
\end{figure}

    We optimized the layout of the LACT to achieve full coverage of the entire LHAASO-KM2A array with the minimum number of telescopes, in order to maximize the utilization of the LHAASO-KM2A detector. ASTRI MINI Array  \citep{vercellone2022astri} has shown that several telescopes with wide separation can have an effective area $\sim 10^6 ~ \rm {m^2}$.  As a preliminary design we used 8 telescopes and the layout of the telescopes is shown in Fig.\ref{fig:layout}. 
The distance between different telescopes is nearly $350~\rm m$. Since we only focus on gamma rays with energies greater than $10~\rm TeV$, we can see from Fig.\ref{fig:photon_density} that the lateral photon density at $350~\rm m$ is nearly $200 -300 ~\rm m^{-2}$. Therefore a $10~\rm TeV$ gamma event can easily be observed simultaneously by more than two telescopes and we can do the stereoscopic reconstruction. At the same time, as the distance increases, the image will get closer to the edge of the camera. Therefore, we need a camera with large fields of view in order to measure the shower accurately enough at such a large distance.


For each telescope, We employed a Davies-Cotton design of mirror to image the cherenkov flash to a SiPM-based camera. The diameter of the telescope is set to be $6~\rm m$, with a  focal length of $8~\rm m$. The camera consists of 2224 pixels with a diameter of $2.58 ~\rm{cm}$ for each pixel. 
{The telescope's field of view(FOV) has a diameter of $9.6^{\circ}$}.  Additionally, we utilized a Winston-cone light collector with a light collection efficiency of 0.93 across all angles. All pixels have been divided into neighboring sectors for trigger purposes. We estimate the night sky background (NSB) using the real data measured at the LHAASO site: {the NSB rate is about $0.1 ~\rm{pe}~ \rm{m^{-2}}\rm{ns^{-1}} \rm {deg^{-2}}$ }, considering here we use an integration time for $100~\rm ns$ and the mirror area is about $28~ \rm {m^2}$, and the pixel size is $\sim 0.19^{\circ}$, so the NSB photoelectrons added in this analysis is  $10~\rm p.e.$ per pixel.

\section{Monte Carlo Simulation and event reconstruction }
\subsection{Simulation setup}
The most important performance of the IACT array is the differential sensitivity for a given exposure time (such as 50 hours) for point sources (Like Crab). The differential sensitivity is limited by either the signal to noise ratio and the number of photons detected. Previous research  \citep{acharya2013introducing} showed that above $10 ~\rm TeV$, the differential sensitivity for point sources is statistic-limited. In this regime, the sensitivity is only determined by the number of photons detected. Thus a better gamma/proton separation power, which can suppress the background and increase the signal to noise ratio, can hardly improve the sensitivity in this energy range. Furthermore, the LHAASO-KM2A array shows a significant improvement of gamma/proton separation ability in comparison with IACTs only above 20 TeV. At first glance, the inclusion of information obtained by LHAASO-KM2A  cannot improve the sensitivity of IACTs. However, as the exposure time T increases, the transition energy $E_t$ above which the observation goes into statistic-limited also increases. Also as mentioned above, the LHAASO array (both LHAASO-KM2A and LHAASO-WCDA) will carry on the sky survey due to their large FOV and nearly full duty cycle, thus LACT is designed to operate in the "staring"(deep observation) mode, which will concentrate on several brightest sources for exposures of more than several hundred hours each year. Furthermore, the latest LHAASO observations  \citep{cao2023first}  \citep{cao2021ultrahigh} reveal that most sources in the Galactic plane are extended. Compared with point sources the extended source should also have a larger 
$E_t$ due to a much higher background rate.  Thus for the extended source with "staring" mode observations, we expect the LHAASO-KM2A gamma/p separation can improve the sensitivity of IACTs. In the following analysis, we consider a disk-like extended source with a radius of $0.5^{\circ}$ and an exposure of 50,200 and 500 hours, respectively.

{\footnotesize 
\begin{table*}[htbp]
	\centering
	\begin{tabularx}{\textwidth}{|X|X|X|X|X|X|X|}
	\hline
	particle type & Energy Range & Zenith angle(deg) & Viewcone angle(deg) & Scatter radius(m) & spectral index & number of events \\ \hline
	Gamma&  $1\rm TeV$ - $300\rm TeV$ & $20^{\circ}$ & $0.5^{\circ}$ & 1000 & -2 & 32 million \\ \hline
	Proton& $1\rm TeV$ - $400\rm TeV$ & $20^{\circ}$ & $2.0^{\circ}$ & 1000 & -2 & 106 million \\ \hline
	\end{tabularx}
	\caption{Table1: simulation parameters}
	\label{tab:para}
\end{table*}
} 

For air shower simulation, We used the CORSIKA  \citep{heck1998corsika}(version 7.6400) to generate the extensive air shower (EAS). Utilizing the Monte Carlo method, this code simulates the generation of a cascade of secondary particles in the atmosphere when a high-energy primary particle enters the Earth's atmosphere and interacts. In this work, we used the FLUKA and QGSJETII models for low-energy and high-energy hadronic interactions respectively and EGS4 for electromagnetic interactions. The observation {altitude} and geomagnetic field were set to be the same as 
LHAASO Sites (Haizi Mountain, Sichuan province,  China). As mentioned above, the improvement brought by the muon detector is important when gamma-ray emission is extended and the direction cut can not effectively suppress the background. So in this work, we only focus on the on-axis performance for extended sources. {The off-axis performance with large offsets (larger than 1.5 degrees) begins to degrade due to worse optical quality. However, in this work, for sources with an extension of about $0.5^{\circ}$, such effects have only a minor impact on our final results. The detailed performance of the array with different offsets will be investigated in future works.} We generated gamma-ray events with energy 
ranging from $1 ~\rm TeV$ to $300 ~\rm TeV$ and a viewcone of $0.5^\circ$. Proton background events should be simulated over a solid angle larger than the FOV of the cameras  \citep{federici2011design}. However, considering the excellent suppression ability of the muon detector and a large field-of-view($\sim 9.6^{\circ}$)
of our telescope, this would require an unbearable amount of CPU time and disk space. Therefore, we decided to simulate the proton events with viewcone of $2.0^{\circ}$ (it is relatively large compared to the gamma source size) and it shouldn't significantly affect the main conclusions, but will be later verified with more detailed MC simulations.
{Gamma ray events will be reweighted to a spectral index of -2.57, and proton events to -2.7.}The scatter radius was set to be 1000\rm~m for both gamma ray events and proton events. The parameters used in the simulation are reported in Table\ref{tab:para}. 
After shower generation, we got two output files: secondary particles information at the observation level and the cherenkov photon data. The secondary particles files are input into the 
G4KM2A \citep{chen2019detector}, the official simulation software of LHAASO-KM2A based on GEANT4, which has been extensively validated on collected data \citep{aharonian2021observation} . The photon files, instead are given as input to the \text{sim\_telarray} simulation package  \citep{bernlohr2008simulation}, which has been specifically developed to simulate telescope response. It includes the ray-tracing on the telescope mirror and the electronic signal generation. For simplicity, we didn't consider the electronic waveform of SiPM here and used the photon arriving at the pixel considering the quantum efficiency of SiPM instead. This simulation used the telescope configuration mentioned in Sec.2.  The trigger logic is to have at least 4 p.e. per pixels, and at least 3 pixels per sector to have a corresponding trigger. This value assumes that the 10 p.e. possion background, are seen as a baseline, is subtracted before applying the trigger criteria.

After the procedure mentioned above, the simulation output file of the two instruments are merged together. For demonstrating the improvement of differential sensitivity, we reconstruct the events individually and use LHAASO-KM2A to do the gamma/proton separation if the events pass the selection criteria of LHAASO-KM2A. 

\subsection{Event reconstruction}
For the reconstruction using IACT individually, in order to exclude the noisy pixels, we first need to do the image cleaning. We used the two-level tail-cuts to accomplish this, which means only pixels containing more photoelectrons than high level (low level) with a neighboring pixel above low level (high level) are kept in the image  \citep{bernlohr2013monte}. Here, two levels are set to be 10 p.e and 20 p.e. 
In the analysis, in order to guarantee the quality of images, we demand that at least two telescopes pass the size cut, and at least one telescope's image passes the edge cut.   Size cut requires that the total photoelectrons in the image must exceed 200 p.e where the edge cut requires that the distance between the camera center and the centroid center of images must be less than $3.5^\circ$.  All telescopes passing the size cut will be used for the direction reconstruction. To reduce the impact of truncation, we only use the telescopes that pass the edge cut for the gamma/proton separation and energy reconstruction.
After image cleaning, 
we can utilize the hillas parametrization  \citep{hillas1985cerenkov} for the nearly elliptical image. The parameters used in this work include the centroid position of the image ($\rm x_{cog}, y_{cog}$), Length (L), Width (W), Size (A), and Alpha ($\alpha$). Length (Width) is the RMS  spread of the major (minor) axis of the ellipse images. Size is the total number of photoelectrons in the 
image after cleaning. The parameter Alpha is the orientation of the major axis. After parameterization, we used the classical major axis intersection method for the direction reconstruction  \citep{bernlohr2013monte}. We map the major axis of each telescope into a reference telescope frame and implement the pair-wise intersection. The weights used for intersection pair is :
\begin{equation}
    \omega_{ij} = \left(\frac{A_i A_j}{A_i + A_j} \right)^2 \cdot sin^2(\phi_{ij}) \delta_i^2 \delta_j^2
\end{equation}

\begin{figure}
  \centering
  \includegraphics[width=\columnwidth]{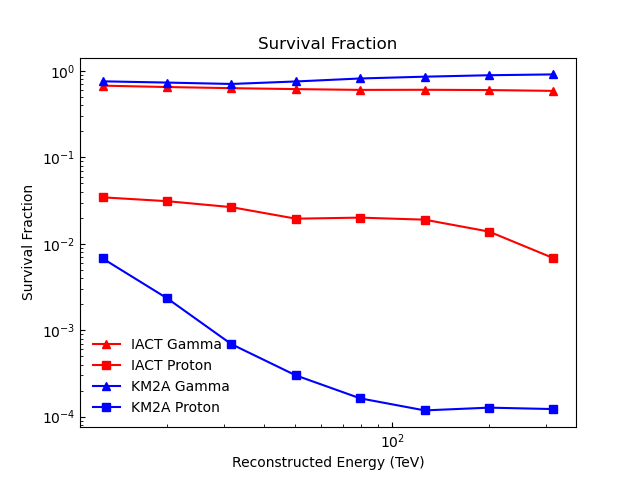} 
  \caption{The survival fraction of gamma ray events and proton events for IACT array and LHAASO-KM2A.}
  \label{fig:survival}
\end{figure}

Here, $A_i, A_j $ is the total photoelectrons (size) in the $i, j$ images and the $\phi_{ij}$ is the angle between the major axis in the reference frame, and $\delta=1 - \frac{W}{L}$  is used because an elongated major axis can provide a more accurate location. For gamma/proton separation, we used the mean reduced scaled length (MRSL) and mean reduced scaled width (MRSW) introduced in \citet{aharonian2006observations}, which is defined below:
$$
\begin{aligned}
    MRSL = \frac{1}{n}\Sigma_{i} \frac{L_i - <L>}{\delta_l} , 
    MRSW = \frac{1}{n}\Sigma_{i}  \frac{W_i - <W>}{\delta_w}
\end{aligned}
$$
$<L>,<W> \text{and } \delta_l, \delta_w$ are obtained from the lookup table.The lookup table was obtained from the simulation of point sources with different offset angles.  Here, we used a shape cuts similar to that used in HESS  \citep{berge2006detailed}, the cut value is set to be $-2.0< MRSL < 2.0, -2.0<MRSW<0.9$. The survival fraction of gamma and proton using this cut is shown in Fig.\ref{fig:survival}. Energy reconstruction is similar to the gamma/proton separation, owing to that different energy showers will have different size given the  different impact parameter.  We used a lookup table of (A/E)  versus the impact parameter and log10(A) to get the estimation of the shower energy. Here, we used the (A/E) for easier interpolation.

\begin{figure}
  \centering
  \includegraphics[width=\columnwidth]{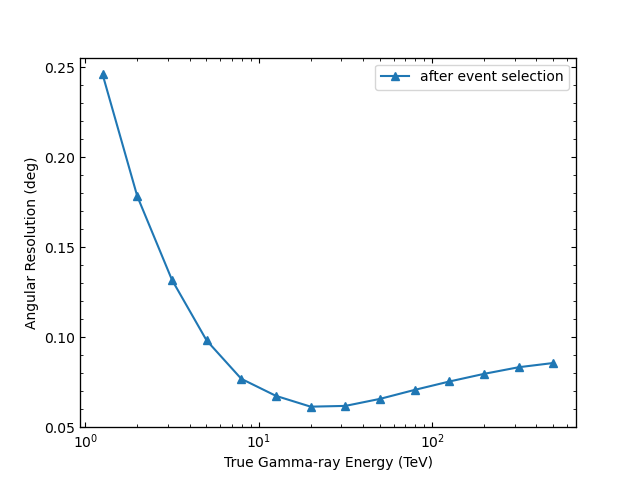} 
  \caption{ Angular resolution of the IACT array.}
  \label{fig:angular}
\end{figure}

\begin{figure}
  \centering
  \includegraphics[width=\columnwidth]{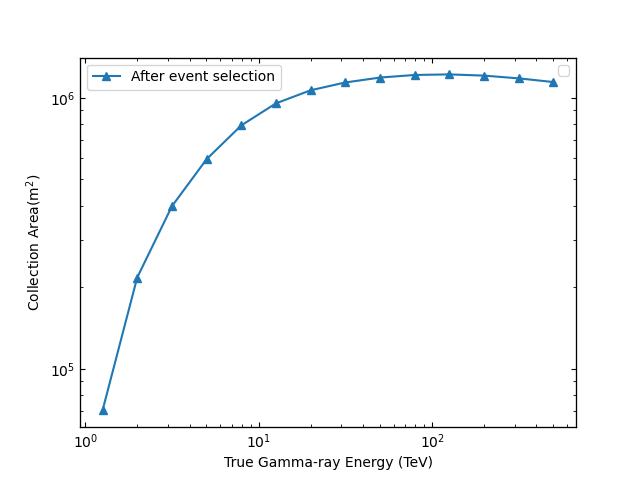} 
  \caption{Collection area of IACT array. It's the area before the shape cut and direction cut. }
  \label{fig:effe_area}
\end{figure}

Fig.\ref{fig:angular} has shown the angular resolution of LACT, {which is defined as the $68\%$ containment radius of the reconstructed event positions from a point-like source.} Although the multiplicity is less due to the larger separation distance, LACT can still provide an excellent direction reconstruction($\sim 0.06^\circ @ 20\rm TeV$) compared to LHAASO-KM2A.  Fig.\ref{fig:effe_area} has shown the collection area (before the gamma/proton separation and the direction cut) of LACT, at $30 \rm~ TeV$ LACT can achieve a collection area larger than $1\rm~{km^2}$, which demonstrate that it can cover the whole LHAASO-KM2A array. The edge cut limits the collection are above $50 \rm~TeV$, as it starts to cut events with large impact parameters.
The energy resolution of the LACT is approximately 15\% - 20\%, with the reconstructed bias being less than 5\%. As the energy exceeds tens of TeV, the energy resolution begins to deteriorate gradually due to the leakage of the image. In the future, employing additional parameters and methods such as random forests could potentially improve the energy resolution and reduce bias.

In the subsequent analysis, taking into account a gamma-ray source size of $0.5^\circ$ and limited PSF, we applied a directional criterion of $d\theta < 0.6^\circ.$ This means that the angular distance in the nominal plane between the center and the reconstructed direction should be less than $0.6^\circ$.

The detailed reconstruction method of LHAASO-KM2A is shown in \cite{aharonian2021observation}. Here, we only focus on the gamma/proton separation in LHAASO-KM2A. Compared to the cosmic ray background events, gamma events are muon-poor and electron-rich. Therefore, we use the R factor (ratio of the number of muons and the number of electrons) to do the gamma/proton separation. R is defined by the following formula:
\begin{equation}
    R=\log \left(\frac{N_\mu+0.0001}{N_{\mathrm{e}}}\right)
\end{equation}
The $N_\mu$ is the number of muon particles at a distance of 15-400 meters from the shower axis, and $N_e$ is the number of electrons at a distance of 0-400 meters from the shower axis.

We apply the following event selection: 1) the number of EDs and the number of particles for the reconstruction (in all EDs) are both greater than 10. 2) the number of particles detected within $40 \rm ~m$ from the shower core is larger than that within $40-100 \rm ~m$. 3) the shower age is between 0.6 and 2.4. 4) the distance between the reconstruction core position and the edge of the array should be less than $20 \rm ~m$. In consideration of the limited effective area of LHAASO-KM2A below 10 TeV, here we only simulate the LHAASO-KM2A response for shower with energy above $10~\rm TeV$.

Fig.\ref{fig:survival} have shown the survival fraction of gamma events and proton events after the cut $ R < -2.35$. We can conclude that compared to the shape cut of the IACT array,  LHAASO-KM2A not only offers a higher gamma event retention ratio but also more effectively eliminates cosmic ray background. 

\begin{figure}
  \centering
  \includegraphics[width=\columnwidth]{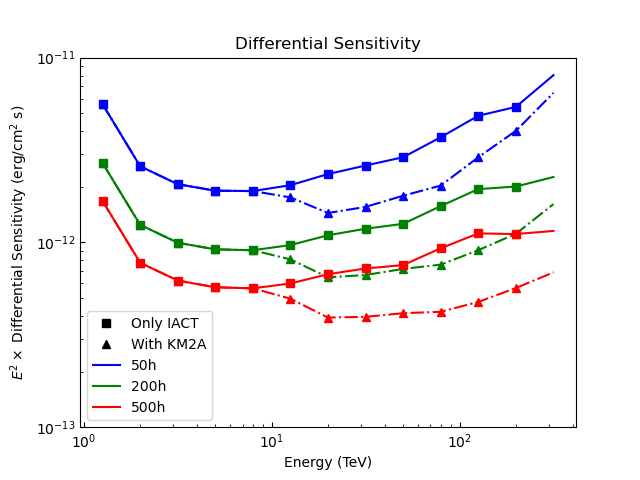} 
  \caption{Differential Sensitivity for 50h, 200h and 500h. The solid line represents the sensitivity of the IACT array, and the dotted line represents the results using the LHAASO-KM2A to do the gamma/proton separation}
  \label{fig: sens}
\end{figure}

Then we calculated the sensitivities for IACTs with and without the information from MDs in LHAASO-KM2A. Differential sensitivity is defined as the minimum flux to obtain a 5-standard-deviation detection. The significance is calculated using the Li-Ma equation  \citep{li1983analysis}.
\begin{equation}
    S_\gamma = \rm \sqrt{2} (N_{on}\ ln(\frac{(1 + \alpha) N_{on}}{ \alpha ( N_{on} + N_{off})}) +N_{off} ln(\frac{(1 + \alpha ) N_{off}}{N_{on} + N_{off}}))^{1/2}
\end{equation}
The alpha in the formula is set to be 0.2 and the energy is divided into five bins per decade. Besides the significance, we also require at least ten detected gamma rays per energy bin.
In the joint analysis with LHAASO-KM2A, if an event has passed the LHAASO-KM2A event preselection, we will use LHAASO-KM2A to do the gamma/proton separation, otherwise, we will use the shape cut mentioned before. In Fig.\ref{fig: sens}, we showed the sensitivities for both cases for 50 hours, 200 hours, and 500 hours. For the source with an extension of $0.5^\circ$,  {we used an extraction region with the radius of $0.6^{\circ}$ to account for the PSF effects,  } and even for 50 hours, the sensitivity in most of the energy ranges is determined by significance. The transition value $E_t$  mentioned before is about $100 ~\rm TeV$, $150 ~\rm TeV$, and more than $300 ~\rm TeV$ for the exposure 50 hours, 200 hours, and 500 hours, respectively. 

With the inclusion of KM2A,  the enhanced gamma-ray retention ratio grants LACT a larger effective area, thus improving its sensitivity at energies above 20 TeV. Notably, within the range of energies less than $E_t$,   the increased cosmic ray rejection rate could, in principle, enhance the signal-to-noise ratio of the LACT, thereby yielding a greater improvement. As shown in Fig.\ref{fig: sens} the sensitivity above 20 TeV   are improved by about $25\% - 60\%$ for all three exposure time we used here.


\section{Disscusion and Conclusions}

In this work, we found that the inclusion of Muon information from LHAASO-KM2A can significantly improve the sensitivity of IACTs on extended sources. For 50 hours exposure the improvement is larger than $20\%$ in the energy range from $20-100~\rm TeV$. And for "staring" mode observation which is the target for LACT, such improvement can be achieved in a much wider energy range (for 500 hours exposure in the energy range $20-300~\rm TeV$).  Such long exposure is not common for current IACTs. But LACT is planned to be built in the LHAASO site to provide a better angular resolution observation for LHAASO detected sources. The sky survey will be accomplished by LHAASO, LACT will focus on only a few of the most interesting sources to provide detailed spatial information. Thus such long exposure observations are feasible. {\bf The observation time per year for WFCTA in the same site of LHAASO can reach 1400 hours, mainly from October to April. Thus, the feasible observational time for LACT can be 200 hours for sources with \( l \sim 30^{\circ} \) and as high as 400 hours per year for sources with \( l \sim 100^{\circ} \), where \( l \) is the galactic longitude. }

Recently, LHAASO has released the first source catalog  \citep{cao2023first}, in which 90 sources are identified and more than 70\% of them are extended even with the limited angular resolution of LHAASO. Remarkably, the brightest UHE sources,1LHAASO J1908+0615u and 1LHAASO J2228+6100u, which are of prime interests in tracing the PeV CR origin in the Milky Way,  are all extended sources with an extension of about $0.5^\circ$. Our studies here revealed that the joint reconstruction by LACT and LHAASO-KM2A are power tool to study these sources, for which LACT can improve to understand the morphology of these regions,  which would be crucial in  unveiling 
the acceleration and propagation mechanism of cosmic rays.

The current joint reconstruction we described here is still very preliminary, we only used the ratio of muon/electron to do the Gamma/Proton separation. But LHAASO-KM2A can provide much more information. Indeed, the measurements from both LHAASO-KM2A and IACTs contain both the lateral and longitudinal distribution information of the air shower and have great potential to study the air shower comprehensively.   For example,  a relatively good core resolution ($\sim 3-4 ~\rm m$) derived from LHAASO-KM2A   can be used to help the IACT array to reconstruct the core position and shower direction simultaneously ( Algorithm 6 in \citet{hofmann1999comparison} ), In this case, the angular resolution of IACTs can be further improved. On the other hand, the IACT array can reconstruct the core position individually for the events outside the LHAASO circle, using the core information provided by the IACT array, we can study the lateral distribution of the particle by using LHAASO-KM2A. This can even extend the effective area of LHAASO-KM2A. 

Although here we only show the performance of eight telescopes, LACT is designed to have 32 telescopes and will have two working modes. In mode 1, LACT will work as 4 subarrays as we showed here, each subarray will observe the different sources thus we can have 4 times more observation time. In mode 2, all 32 telescopes will work as a large array. Due to the reduced spacing between the telescopes, the multiplicity can be significantly greater than 2.  Compared to Mode 1, this configuration yields enhanced angular resolution and a lower energy threshold. In this mode, LACT would become the most sensitive IACT array in the Northern Hemisphere. In conclusion, the joint analysis between IACTs and EAS arrays has great potential to improve both the effective area and angular resolution for current ground based gamma-ray astronomy, which would advance our understanding for the extreme high-energy phenomena of the non-thermal universe.
\section*{Acknowledgements}
Rui-zhi Yang is supported by the NSFC under grants 12041305 and the national youth thousand talents program in China. This work is also supported by Sichuan Science and Technology 
Department, Institute of High Energy Physics through the
grants of 2023YFSY0014 and E25156U1, respectively. This study is also supported by the following grants: the Sichuan Province Science Foundation for Distin-guished Young Scholars under grant No. 2022JDJQ0043; the Sichuan Science and Technology Department under grant No. 2023YFSY0014; the Xiejialin Foundation of IHEP under grant No. E2546IU2; the National Natural Science Foundation of China under grants No. 12261141691; the Innovation Project of IHEP under grant No. E25451U2.


\bibliographystyle{elsarticle-harv} 
\bibliography{biblio.bib}






\end{document}